\documentclass[11pt]{article}
\usepackage{amsfonts}
\usepackage{amsmath}
\usepackage{amsthm}
\usepackage[font={small,it,stretch=1.2}]{caption}
\usepackage{changepage}
\usepackage{commath}
\usepackage{enumerate}
\usepackage[margin=1.0in]{geometry}
\usepackage{graphicx}
\usepackage{hyperref}
\usepackage[utf8]{inputenc}
\usepackage{mathtools}
\usepackage[section]{placeins}
\usepackage{natbib}
\usepackage{setspace}
\usepackage{subcaption}
\usepackage{titling}
\usepackage[table]{xcolor}
\setcitestyle{round}
\hypersetup{
    colorlinks=TRUE,
    linktoc=all,
    linkcolor=violet,
    citecolor=teal
}
\newcommand{\LG}{\cellcolor{lightgray}}

\AtBeginDocument{
    \setlength\abovedisplayskip{-0.1cm}
    \setlength\abovedisplayshortskip{-0.1cm}
    \setlength\belowdisplayskip{0.4cm}
    \setlength\belowdisplayshortskip{0.4cm}
}

\title{Factors affecting power in stepped wedge trials when the treatment effect varies with time}
\author{
Avi Kenny$^{1,2,*}$, Emily C. Voldal$^3$, Fan Xia$^4$,\\
Kwun Chuen Gary Chan$^5$, Patrick J. Heagerty$^5$, James P. Hughes$^5$ \\ \\
$^1$Department of Biostatistics \& Bioinformatics, Duke University\\
$^2$Global Health Institute, Duke University\\
$^3$Vaccine and Infectious Disease Division, Fred Hutch Cancer Center\\
$^4$Department of Epidemiology and Biostatistics, University of California, San Francisco\\
$^5$Department of Biostatistics, University of Washington\\
* avi.kenny@duke.edu
}

\begin{document}

\maketitle

\begin{center}
    \section*{Abstract}
\end{center}

\begin{adjustwidth}{0.5in}{0.5in}
    \textbf{Background}\\
    Stepped wedge cluster randomized trials (SW-CRTs) have historically been analyzed using immediate treatment (IT) models, which assume the effect of the treatment is immediate after treatment initiation and subsequently remains constant over time. However, recent research has shown that this assumption can lead to severely misleading results if treatment effects vary with exposure time, \emph{i.e.} time since the intervention started. Models that account for time-varying treatment effects, such as the exposure time indicator (ETI) model, allow researchers to target estimands such as the time-averaged treatment effect (TATE) over an interval of exposure time, or the point treatment effect (PTE) representing a treatment contrast at one time point. However, this increased flexibility results in reduced power.\\
    
    \noindent\textbf{Methods}\\
    In this paper, we use public power calculation software and simulation to characterize factors affecting SW-CRT power. Key elements include choice of estimand, study design considerations, and analysis model selection.\\
    
    \noindent\textbf{Results}\\
    For common SW-CRT designs, the sample size (clusters per sequence or individuals per cluster-period) must be increased substantially, commonly by a factor of 1.5 to 3, but often by much more, to maintain 90\% power when switching from an IT model to an ETI model (targeting the TATE over the study). However, the inflation factor is lower for TATE estimands over shorter periods that exclude longer exposure times. In general, SW-CRT designs (including the ``staircase'' variant) have much greater power for estimating ``short-term effects'' relative to ``long-term effects''. For an ETI model targeting a TATE estimand, substantial power can be gained by adding time points to the start of the study or increasing baseline sample size, but surprisingly little power is gained from adding time points to the end of the study. More restrictive choices for modeling the exposure time or calendar time trends (e.g., splines or linear terms) have little effect on power for TATE estimands but increases power for PTE estimands. If the effect curve is constant after a washout period, a ``delayed constant treatment'' model that uses exposure time indicators during the washout period but assumes a constant effect thereafter can slightly increase power relative to an IT model that discards washout period data.\\ \\
    
    \noindent \textit{Keywords}: stepped wedge, cluster randomized trial, staircase, power, sample size, time varying treatment effects, treatment effect heterogeneity
\end{adjustwidth}

\doublespacing

\section{Background}\label{sec_intro}

\begin{figure}[ht!]
    \centering
    \includegraphics[width=0.8\linewidth]{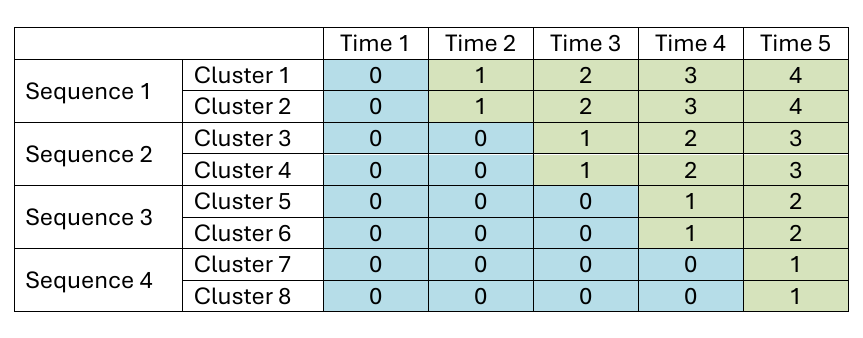}
    \captionsetup{width=0.9\linewidth}
    \caption{Exposure time in a typical stepped wedge study. The labels in the column headers represent calendar time and the numbers in each cluster-period cell represent exposure time. A blue cell containing a 0 represents a cluster in the control state and a green cell containing a number greater than zero represents a cluster in the treatment state.}
    \label{sw_diagram}
\end{figure}

The stepped wedge cluster randomized trial (SW-CRT) is a popular study design in which clusters of individuals are randomized to an intervention in a phased rollout manner, such that all clusters eventually receive the intervention \citep{hemming2015stepped}. Data from SW-CRTs have historically been analyzed using immediate treatment (IT) models which represent a large class of statistical models that assume the effect of the treatment is achieved immediately after the start of intervention and that it remains constant over time since initiation. However, recent research has shown that making this assumption can lead to severely misleading results if the treatment effect varies with exposure time which is defined for a given cluster as the amount of time that has passed since that cluster crossed over from the control state to the intervention state \citep{kenny2022analysis}; see Figure \ref{sw_diagram} for a schematic diagram of a typical stepped wedge study illustrating the concept of exposure time.

Time-varying treatment effects are theoretically plausible for many interventions that are expected to take time for the full effect to be reached, such as weight loss or behavior change interventions. They are also empirically observable in applied studies. For example, in a reanalysis of a stepped wedge trial examining the impact of the disinvestment of weekend health services from twelve hospital wards in Australia (\citealp{haines2017impact}; Trial 1), there was evidence that the treatment effect increased substantially at the third exposure time point and beyond \citep{kenny2022analysis}.

The large differences in power for estimating short-term and long-term estimands have important consequences for study design. Long-term estimands are typically prioritized whenever the main scientific goal is to understand the sustained impact of an intervention beyond its initial implementation phase. Researchers interested in long-term estimands may want to consider the feasibility of using a parallel-arm design, as this will typically lead to much greater power. If a stepped wedge design is used, it may sometimes be useful to collect additional outcome data after all clusters are in the treatment state to understand how outcome levels change over time, even if no control group is available. If short-term and long-term estimands are equally appealing, the feasibility of estimating short-term effects may weigh into the decision of which estimand to target with the primary analysis.

Several models have been proposed that allow for time-varying treatment effects in SW-CRT analysis \citep{kenny2022analysis,maleyeff2023assessing}, but the increase in model flexibility required to target time-varying treatment effect estimands comes at the cost of reduced statistical power, which is a concern for researchers designing, planning, and analyzing SW-CRTs. Specifically, these models must account for the fact that the functional form of the \textit{effect curve} (the treatment effect as a function of exposure time) is unknown, but ultimately allow researchers to target a specific estimand. In this context, an estimand is a summary or feature of the effect curve; common estimands include the average value of the curve over the course of the study and the value of the curve at a specific point in time. Denoting a value of the curve at a specific point in time as the estimand of interest is similar to what is often done in parallel-arm cluster trials, in which researchers specify in advance that they are interested in, for example, the difference in the outcome between arms at month 6. Denoting the average value of the curve as the estimand of interest may be done if there is no specific time point of particular interest. As we will see, the choice of estimand will substantially affect power.

One model used when treatment effects vary with time is the exposure time indicator (ETI) model, studied by \cite{kenny2022analysis}, which includes indicator variables corresponding to specific exposure times and makes no assumptions about the shape of the effect curve. \cite{hughes2024sample} derived an analytic variance formula for treatment effect estimators in an ETI mixed effects model that can be expressed as linear combinations of the time-specific treatment effect parameters. They used this formula to show how power can be estimated and demonstrated that under an identity link function, power only depends on the value of the summary estimand (e.g., the average value of the curve) and does not depend on the shape of the curve. Subsequently, at least two power calculation software packages have implemented this formula, including the \texttt{swCRTdesign} R package \citep{voldal2020swcrtdesign} and the NIH Research Methods Resources stepped wedge sample size calculator \citep{murray2024design}. However, no study has systematically examined factors affecting power in SW-CRTs when the treatment effect varies with exposure time. In this paper, we use power calculation software and simulation to do so, examining factors related to (1) estimands of interest, (2) study design, and (3) modeling choices. Specifically, we make the following contributions:

\begin{enumerate}
  \item For a classic stepped wedge design involving a correctly-specified IT model, we compute the relative increase in sample size necessary to use an ETI model for different choices of estimand.
  \item For settings in which there may be time-varying treatment effects, we characterize the power of an ETI model for targeting different estimands of interest.
  \item We determine the effects of study design choices, including the addition of data collection time points at the start or end of the study and the use of the ``staircase'' variant of the standard SW-CRT \citep{grantham2024staircase}, on power and ability to effectively target different estimands.
  \item We examine the impact of different modeling choices on power, including more restrictive choices for the calendar time or exposure time trends and use of a ``delayed constant treatment'' model that assumes a constant treatment effect after a washout period.
\end{enumerate}

The organization of the remainder of this paper is as follows. In Section \ref{sec_methods}, we introduce estimands and models, outline simulation methods, and describe how we utilize power software. In Section \ref{sec_results}, we describe results related to the four bullet points above. In Section \ref{sec_discussion}, we discuss practical implications of these results on the design and analysis of SW-CRTs.

\section{Methods}\label{sec_methods}

\subsection{Estimands}\label{ssec_estimands}

Suppose $Y_{ijk}$ represents the outcome of interest in cluster $i\in(1,2,...,I)$ at time point $j\in(1,2,...,J)$ for individual $k\in(1,2,...,K_{ij})$, where $K_{ij}$ is the number of individuals observed in cluster $i$ at time $j$. We restrict attention to the case of cross-sectional data with continuous outcomes and assume that data are generated from a mechanism with the following mean model, which implicitly conditions on the design matrix:

\begin{equation*}
    E(Y_{ijk}) = \Gamma(j) + \delta(s_{ij}) \,,
\end{equation*}

\noindent where $\Gamma(j)$ is a generic term representing the time trend at time $j$, $s_{ij}$ represents the exposure time of cluster $i$ at time $j$, and $\delta$ is an arbitrary function (subject to the constraint that $\delta(0)=0$) representing the effect curve. This model allows the treatment effect to vary as a function of exposure time, and thus treatment effect summaries can be expressed as functionals of the effect curve $s \mapsto \delta(s)$. In the context of this mean model, the \textit{point treatment effect} (PTE) at exposure time $s_1$ is defined as $\text{PTE}(s_1) \equiv \delta(s_1)$ and the \textit{time-averaged treatment effect} (TATE) between exposure times $s_1$ and $s_2$ is defined as

\begin{equation*}
    \text{TATE}(s_1,s_2) \equiv \frac{1}{s_2-s_1} \int_{s_1}^{s_2} \delta(s)ds \,,
\end{equation*}

\noindent and can be interpreted as the average value of the effect curve over the interval $[s_1,s_2]$. See \cite{wang2024achieve} for a discussion of when these statistical estimands will have a valid causal interpretation. Note that some authors define $\text{TATE}(s_1,s_2) \equiv \frac{1}{s_2-s_1}(\delta_{s_1}+\delta_{s_1+1}+...+\delta_{s_2})$, where the $\delta_s$ terms represent the parameters of an ETI model; we avoid doing so to avoid having the estimand definition depend on the idiosyncracies of a particular design (e.g., the period lengths). Also, our definition is valid regardless of whether observations are indexed by a grid of discrete time points (as is common in the analysis of stepped wedge trials) or indexed in continuous time, and also valid in designs with varying period lengths.

\subsection{Analysis models}\label{ssec_models}

In this section, we briefly define the mixed models that will be considered in this work, all of which model the correlation structure using two random intercept terms, one corresponding to the cluster and one corresponding to the cluster-period, as suggested in \cite{hooper2016sample} and \cite{girling2016statistical}. For a data structure involving a continuous outcome $Y_{ijk}$, the \textit{immediate treatment} (IT) model is given by:

\begin{equation}\label{model_it}
    Y_{ijk} = \Gamma(j) + \delta X_{ij} + \alpha_i + \xi_{ij} + \epsilon_{ijk} \,,
\end{equation}

\noindent where $X_{ij}$ is an indicator that equals one if cluster $i$ is in the treatment state at time $j$, $\delta$ is the corresponding treatment effect scalar parameter, $\Gamma(j)$ is a generic term modeling the calendar time trend at time $j$, $\alpha_i \sim N(0,\tau^2)$ is a random cluster intercept, $\xi_{ij} \sim N(0,\gamma^2)$ is a random cluster-by-time intercept, and $\epsilon_{ijk} \sim N(0,\sigma^2)$ is a model residual. In this paper, we consider the use of categorical time effects (i.e., setting $\Gamma(j)=\beta_j$, such that there is one time trend parameter per discrete time point, as in \citealp{hussey2007design}) and the use of a linear time trend (i.e., setting $\Gamma(j) = \beta_0 + j \beta_1$). The key assumption of the IT model is that the true effect curve $s \mapsto \delta(s)$ is constant for $s>0$, and therefore all estimands considered in this work are equivalent if this model is correctly specified. Also, we define the intraclass correlation coefficient (ICC) to equal $(\tau^2+\gamma^2)/(\tau^2+\gamma^2+\sigma^2)$ and the cluster autocorrelation coefficient (CAC) to equal $\tau^2/(\tau^2+\gamma^2)$; note that some authors refer to the ICC as the ``within-period ICC''.

Next, the \textit{exposure time indicator} ETI model is given by

\begin{equation}
    Y_{ijk} = \Gamma(j) + \sum_{s=1}^S \delta_s I(s_{ij}=s) + \alpha_i + \xi_{ij} + \epsilon_{ijk} \,,
\end{equation}

\noindent where $s_{ij}$ represents the exposure time of cluster $i$ at time $j$ and $S$ is the maximum observed exposure time (e.g., in a standard design, $S=J-1$). This model involves a vector of treatment effect parameters $(\delta_1,\delta_2,...)$, where each parameter $\delta_s$ corresponds to a distinct point treatment effect $\delta(s)$, as defined in section \ref{ssec_estimands}.

In some situations, one may wish to use a model that assumes that the ultimate effect of the treatment is not fully realized for only a subset of the total exposure time of the study. For example, it may be assumed that following implementation, there is some ``ramp-up'' of the treatment effect for one or more time periods, after which the effect of the treatment reaches and remains at a certain level. The time corresponding to this ramp-up is often referred to as a ``washout period'' or ``implementation period'', and historically it has been common practice to either not collect data during the washout period or to discard these data in the analysis stage \citep{caille2024practical}. However, if data are available, an alternative approach for these settings is to use the \textit{delayed constant treatment} (DCT) model given by

\begin{equation}
    Y_{ijk} = \Gamma(j) + \sum_{s=1}^w \delta_s I(s_{ij}=s) + \delta I(s_{ij}>w) + \alpha_i + \xi_{ij} + \epsilon_{ijk} \,,
\end{equation}

\noindent where $w$ is the number of washout periods, chosen in advance based on contextual knowledge. In many applications we would expect $\delta_s$ to be smaller than the final $\delta$. The DCT model can be thought of as a hybrid between the IT and ETI models, allowing the treatment effect to vary arbitrarily for exposure times $(1,2,...,w)$ but assuming a constant treatment effect for exposure times $(w+1,w+2,...,J-1)$. In a typical use case for this model, the constant treatment effect parameter $\delta$ will be of primary interest, whereas the ramp-up parameters $(\delta_1,\delta_2,...,\delta_w)$ are either nuisance parameters or are of secondary interest.

Finally, the \textit{natural cubic spline} (NCS) model with $d$ degrees of freedom is given by

\begin{equation}
    Y_{ijk} = \Gamma(j) + \sum_{s=1}^{d} \delta_s b_s(s_{ij}) X_{ij} + \alpha_i + \xi_{ij} + \epsilon_{ijk} \,,
\end{equation}

\noindent where $(b_1,b_2,...,b_d)$ is a $d$-dimensional natural cubic spline basis \cite{hastie2009elements} and $(\delta_1,\delta_2,...,\delta_d)$ is the corresponding parameter vector. This model is useful when one wishes to limit the total number of model parameters corresponding to the treatment effect structure, since the number of degrees of freedom $d$ is set by the researcher. Note that it is typically a good choice to include an intercept in the construction of the spline basis to allow for the effect curve to have a discontinuity at zero, as this implies that the IT model is a submodel. Of course, other spline bases can be used, such as polynomial splines or linear splines.

\subsection{Using power formulas to estimate sample size ratios}\label{ssec_methods_formula}

We used the R package \texttt{swCRTdesign} \citep{voldal2020swcrtdesign} version 4.0, as it allows for power to be calculated under both IT and ETI mixed models \citep{hughes2024sample}. As of the version 4.1 update in September 2025, this package can calculate sample size directly (in terms of the number of individuals per cluster-period) as a function of desired power; however, we used a simple iterative wrapper algorithm to do so by minimizing the difference between the desired power and the estimated power as a function of sample size. This, in turn, can be run to estimate the sample size ratio (SSR), defined as the sample size required to achieve $\ge 90\%$ power using an ETI model divided by the sample size required to achieve $\ge 90\%$ power using an IT model, holding all other design and data-generating variables fixed and assuming an immediate treatment effect. The ``sample size'' in the SSR can be defined either in terms of the number of individuals per cluster-period (while holding the total number of clusters constant) or the number of clusters per sequence (while holding the number of individuals per cluster-period constant). Importantly, for a given design, the SSR will differ depending on the estimand of interest.

We considered scenarios with a residual standard deviation of 1 and an immediate treatment effect size of 0.2, representing a small effect size on the Cohen's \textit{d} scale \citep{goulet2018review}. ICC values ranged from 0 to 0.15 and CAC values ranged from 0.5 to 1, representing reasonable ranges of these values that might be seen in practice based on the work of \cite{korevaar2021intra}. For some analyses, the ICC and CAC were held constant at 0.05 and 0.75, respectively, roughly equal to the median values observed by \cite{korevaar2021intra}. All designs used were ``balanced and complete'', meaning that each sequence has the same number of clusters, each cluster-period has the same number of individuals, and all clusters are observed at all time period. We considered several designs involving different numbers of sequences, clusters per sequence, and individuals per cluster-period to cover several scenarios commonly seen in practice.

\subsection{Simulation methods}\label{ssec_methods_simulation}

We also conducted a simulation study to evaluate the effects of different modeling choices on statistical power. One set of simulations examines the effect of different time trend parameterizations on power, including an ETI model with a categorical time trend, an ETI model with a linear calendar time trend, and an NCS model with four degrees of freedom and a categorical calendar time trend, all described in section \ref{ssec_models}. A second set of simulations focuses on settings in which there is a washout period and compares a modified IT model that drops data corresponding to the washout period, an ETI model, and the delayed constant treatment (DCT) model described in section \ref{ssec_models}.

Data were generated according to either the immediate treatment model given in \eqref{model_it} with an immediate treatment effect value of $\delta=0.2$ or a model in which the effect curve had an immediate jump from 0 to 0.1 and a subsequent linear increase from 0.1 to 0.3 over the course of the study; in both cases, the TATE over the course of the study is 0.2. The calendar time trend was a linear function that increased from 0 to 1 over the course of the study. A residual standard deviation of 1 was used, the ICC was either 0.01 or 0.1, the CAC was 0.75, and designs used 4 clusters per sequence, 5 individuals per cluster-period, and a varying number of sequences.

Data were analyzed using the \texttt{steppedwedge} R package version 1.0.0 \citep{kenny_sw_2025}, available on CRAN, which allows for estimation of various treatment effect parameters using either a mixed model or the generalized estimating equations (GEE) framework and implements all models described in section \ref{ssec_models}. All simulations were run in R version 4.3.2 and structured using the \texttt{SimEngine} package version 1.4.0 \citep{kenny2025simengine}; code to reproduce all analyses and simulations is available at \url{https://github.com/Avi-Kenny/SW-Power}.

\section{Results}\label{sec_results}

\subsection{Effects of estimand choices}\label{ssec_res_estimands}

\begin{figure}[ht!]
    \centering
    \includegraphics[width=0.9\linewidth]{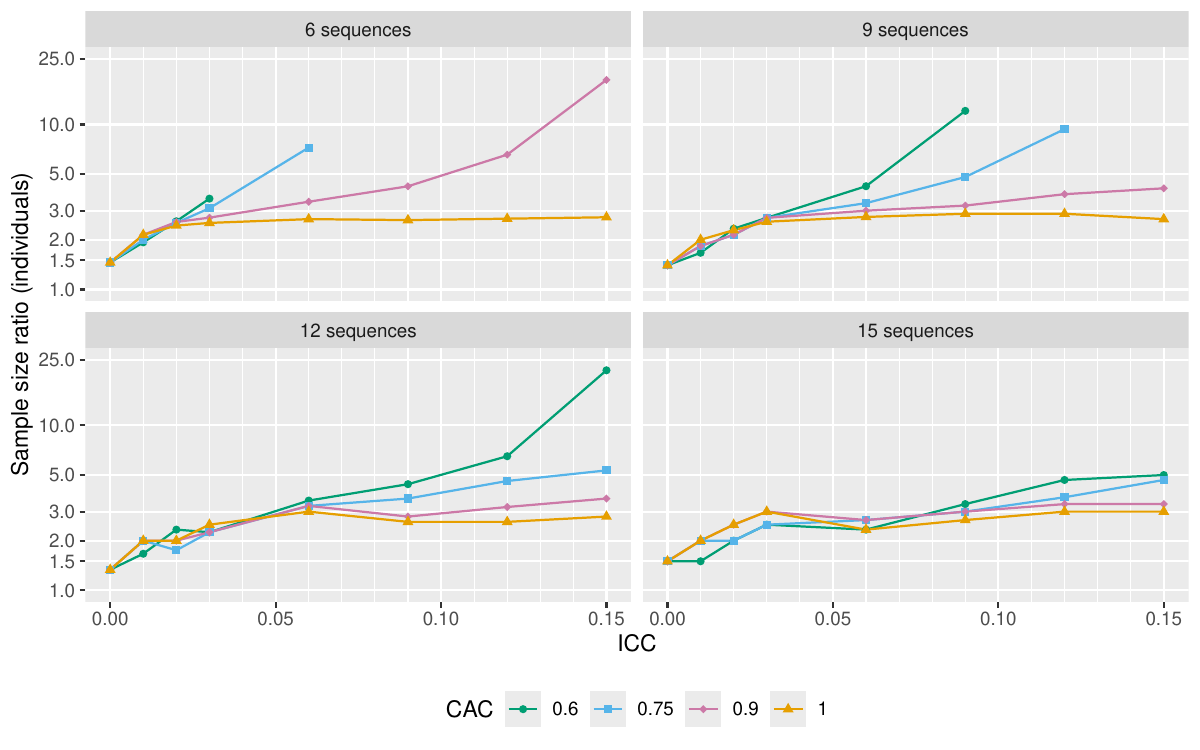}
    \captionsetup{width=0.9\linewidth}
    \caption{Sample size ratio (individuals) required for 90\% power. The line color and point symbols correspond to different CAC values. Power calculations assume data are generated from an IT model with a (standardized) effect size of 0.2 and 4 clusters per sequence. If no point is plotted for a given combination of ICC and CAC then it is impossible to achieve 90\% power for the ETI model by increasing the number of individuals per cluster-period. The Y-axis is displayed on the log scale.}
    \label{fig_SSR_basic_ind}
\end{figure}

To begin, it is useful to consider settings in which the immediate treatment model is correct. In these settings, all estimands are equivalent; that is, $\text{TATE}(s_1,s_2) = \text{PTE}(s_3)$ for all $(s_1,s_2,s_3)$. However, when using an ETI model, it is still necessary to specify which estimand we are targeting so that a corresponding estimator can be chosen. Here, we choose to use an estimator based on the TATE over the course of the study, $\text{TATE}(0,S)$, which is equivalent to the average of the ETI parameter estimators $(\hat{\delta}_1, \hat{\delta}_2, ..., \hat{\delta}_S)$. Figure \ref{fig_SSR_basic_ind} shows the \textit{individual} sample size ratio (SSR), defined as the relative increase in sample size (in terms of number of individuals per cluster-period) necessary to achieve 90\% power when switching from an IT model to an ETI model, both with categorical time effects. The four panels of Figure \ref{fig_SSR_basic_ind} display the individual SSR as a function of the ICC for four designs that vary in terms of the number of sequences. The line color and point symbols correspond to different CAC values.

We see that the individual sample size ratio varies enormously across scenarios from about 1.5 in cases with a low ICC to a factor of 5 or more in cases with higher ICC values and lower CAC values. In some cases, it is not possible to recover 90\% power when switching from an IT model to an ETI model by increasing the number of individuals per cluster alone; this is because the variance of the ETI treatment effect estimator is influenced by the between-cluster variance induced by the random effects, which will not reduce to zero unless the number of clusters is increased, particularly in scenarios with higher ICC values and CAC $<$ 1. Also, the lack of smoothness in the trends is due to discreteness in the number of individuals per cluster-period.

\begin{figure}[ht!]
    \centering
    \includegraphics[width=0.9\linewidth]{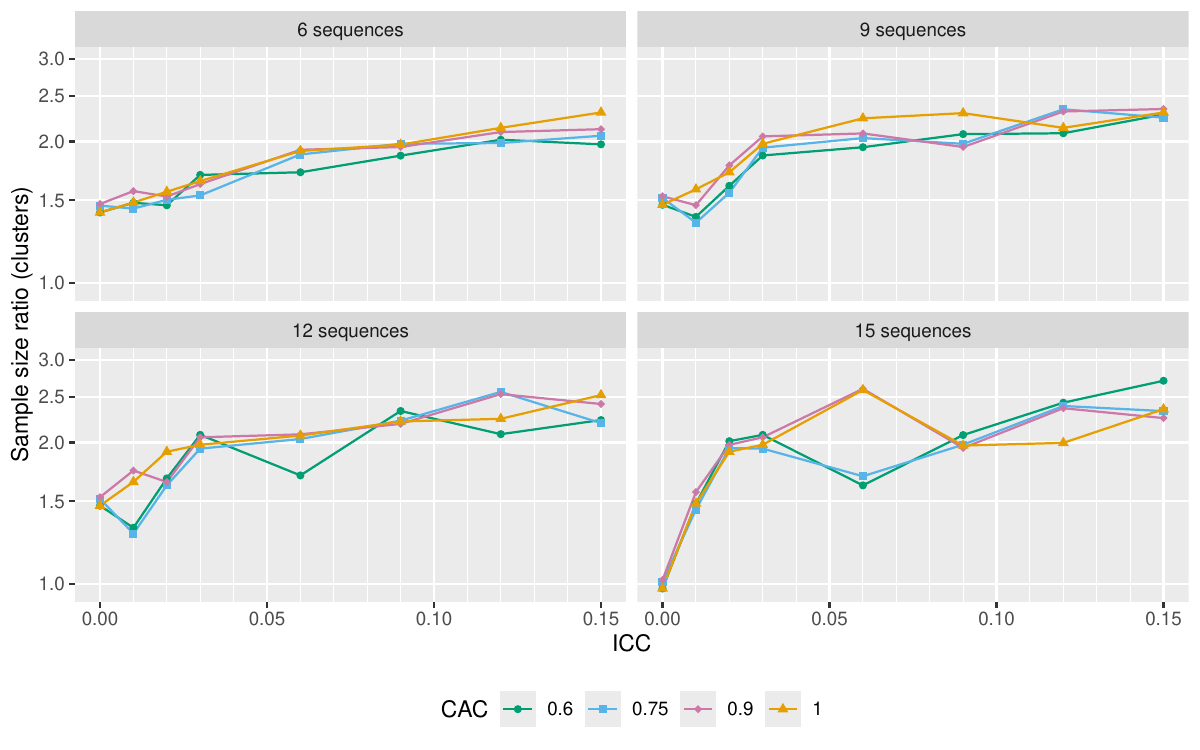}
    \captionsetup{width=0.9\linewidth}
    \caption{Sample size ratio (clusters) required for $\ge$90\% power. The line color and point symbols correspond to different CAC values. Power calculations assume data are generated from an IT model with a (standardized) effect size of 0.2 and 5 individuals per cluster-period. Lack of smoothness in the trends is due to discreteness in the number of clusters. The Y-axis is displayed on the log scale.}
    \label{fig_SSR_basic_clust}
\end{figure}

Given the limitations of the individual SSR, it is useful to consider the \textit{cluster} SSR, defined as the relative increase in sample size (in terms of the number of clusters per sequence) necessary to achieve 90\% power when switching from an IT model to an ETI model, both with categorical time effects. Results are displayed in Figure \ref{fig_SSR_basic_clust}.

We observe that the sample size (number of clusters) often has to be increased by a factor of 1.5 to 2.5 to recover 90\% power when switching from an IT model to an ETI model. Although this is still substantial, these results are not as extreme as the results in Figure \ref{fig_SSR_basic_ind}. Again, the lack of smoothness in the trends is due to discreteness in the number of clusters, since adding one cluster per sequence to a design often involves a large jump in power, leading to a ``rounding'' issue that is not easily resolved. Alternative definitions for the cluster SSR (e.g., involving addition of one cluster at a time to a random sequence in the design) may lead to improved smoothness but are not considered here.

\begin{figure}[ht!]
    \centering
    \includegraphics[width=0.9\linewidth]{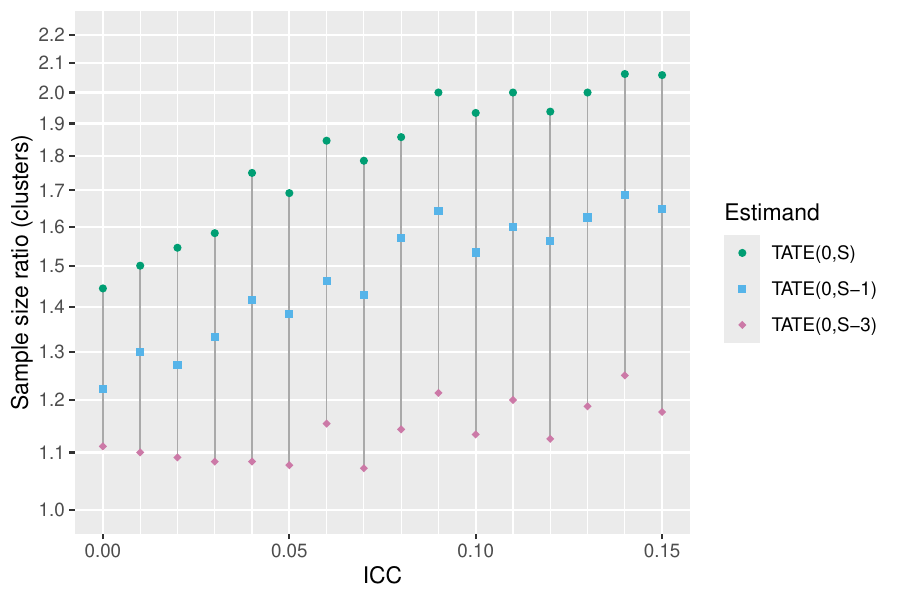}
    \captionsetup{width=0.9\linewidth}
    \caption{Sample size ratio (clusters) required for $\ge$90\% power, shown for three TATE estimands, where $S$ is the total number of sequences in the study. Power calculations assume data are generated from an IT model in a six-sequence design with a (standardized) effect size of 0.2, 5 individuals per cluster-period, and a CAC of 0.75. The Y-axis is displayed on the log scale.}
    \label{fig_SSR_n_omit}
\end{figure}

Next, we examine sample size requirements for different target estimands. We first look at how the cluster SSR changes if, instead of looking at the TATE over the course of the study, $\text{TATE}(0,S)$, we instead look at the ``short-term TATE'', specifically $\text{TATE}(0,S-k)$ for some $k>0$ (where we recall that $S$ is the total number of sequences). That is, we omit $k$ exposure time periods from the end of the estimand definition. Note that we are still assuming that the IT model is correct. Results for a six-sequence design are shown in Figure \ref{fig_SSR_n_omit}.

We see that the cluster SSR lowers considerably as we decrease the number of exposure times over which the TATE is defined; analogous results (not displayed) hold for the individual SSR and for designs with different numbers of sequences. For a six-sequence design with five individuals per cluster-period, a standardized effect size of 0.2, an ICC of 0.05, and a CAC of 0.75, the cluster SSR for estimating $\text{TATE}(0,6)$ is 1.7, but goes down to about 1.4 for estimating $\text{TATE}(0,5)$ and goes down further to just 1.1 for estimating $\text{TATE}(0,3)$. It may feel somewhat counterintuitive that for a given design, we need a much larger sample to estimate $\text{TATE}(0,S)$ than we need to estimate $\text{TATE}(0,S-3)$; this phenomenon occurs because in general, there is far less information in a standard stepped wedge design about point treatment effects corresponding to higher exposure times relative to those corresponding to lower exposure times.

\begin{figure}[ht!]
    \centering
    \includegraphics[width=0.7\linewidth]{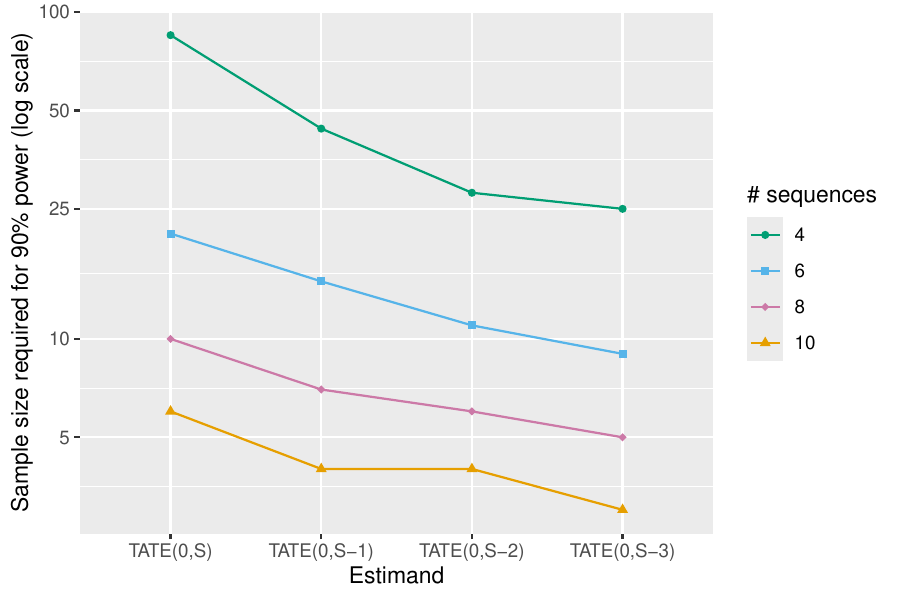}
    \captionsetup{width=0.9\linewidth}
    \caption{Sample size required (number of individuals per cluster-period) for 90\% power with an ETI model as a function of several ``short-term TATE'' estimands of interest. Results correspond to a design with 4, 6, 8, or 10 sequences, 5 individuals per cluster-period, a (standardized) effect size of 0.2, an ICC of 0.05, and a CAC of 0.75. Power calculations assume data are analyzed using an ETI model. The Y-axis is displayed on the log scale.}
    \label{fig_SS_ETI_n_omit}
\end{figure}

Note that, for a given vertical line in Figure \ref{fig_SSR_n_omit}, the IT model we are comparing to is exactly the same for all three points. In some sense, this is not a ``fair'' comparison, since the IT model assumes that the treatment effect is the same for (and uses data from) all exposure times in the design, not just those corresponding to the estimand of interest. We choose to show this comparison in order to highlight the fact that the SSR is highly dependent on the choice of estimand, and targeting a short-term TATE with an ETI model in a standard stepped wedge design can be done with a much smaller sample size relative to what is required to target $\text{TATE}(0,S)$ for a given level of power. However, instead of focusing on the SSR, it may be more intuitive to consider settings in which we allow the treatment effect to vary with exposure time and examine the sample size required (number of individuals per cluster-period) for 90\% power when using an ETI model for different short-term TATE estimands; this is shown in Figure \ref{fig_SS_ETI_n_omit}.

The results of figure \ref{fig_SS_ETI_n_omit} illustrate that, for a given design, the sample size required for achieving 90\% power with an ETI model decreases for estimands $\text{TATE}(0,S-k)$ with greater values of $k$. For example, in an eight-sequence design, we need a sample size of about 10 individuals per cluster-period to target $\text{TATE}(0,8)$ but only about 5 individuals per cluster-period (half the sample size) to target $\text{TATE}(0,5)$. Results are qualitatively similar for other choices of ICC and CAC.

\begin{figure}[ht!]
    \centering
    \includegraphics[width=0.7\linewidth]{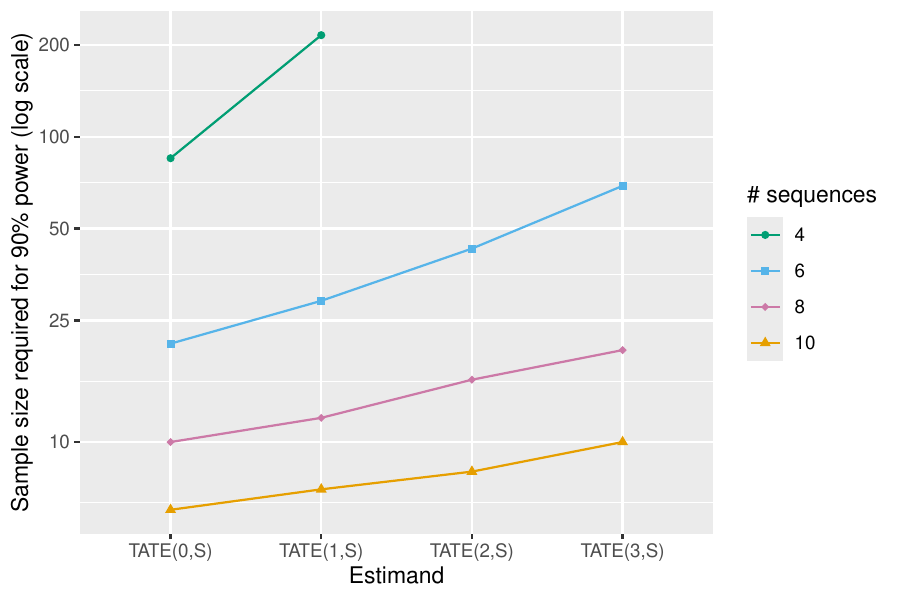}
    \captionsetup{width=0.9\linewidth}
    \caption{Sample size required (number of individuals per cluster-period) for 90\% power with an ETI model as a function of several ``long-term TATE'' estimands of interest. Results correspond to a design with 4, 6, 8, or 10 sequences, 5 individuals per cluster-period, a (standardized) effect size of 0.2, an ICC of 0.05, and a CAC of 0.75. Power calculations assume data are analyzed using an ETI model. If no point is plotted for a given combination of estimand and number of sequences then it is impossible to achieve 90\% power for the ETI model with any sample size in that scenario. The Y-axis is displayed on the log scale.}
    \label{fig_SS_ETI_n_wash}
\end{figure}

Conversely, for a given design analyzed with an ETI model, we would expect that a greater sample size would be needed to target the ``long-term TATE'', $\text{TATE}(k,S)$ (which can be thought of as the TATE following a washout period of length $k$), relative to the sample size necessary to target $\text{TATE}(0,S)$. This is indeed the case; Figure \ref{fig_SS_ETI_n_wash} displays results.

As expected, Figure \ref{fig_SS_ETI_n_wash} shows that for a given design, the sample size required for achieving 90\% power with an ETI model increases for estimands $\text{TATE}(k,S)$ as $k$ increases. For example, in an eight-sequence design, we need a sample size of about 10 individuals per cluster-period to target $\text{TATE}(0,8)$ and a sample size of roughly 20 individuals per cluster-period (twice the sample size) to target $\text{TATE}(3,8)$. In some cases, it is not possible to achieve 90\% power no matter how large the number of individuals per cluster (as discussed above). For example, $\text{TATE}(2,4)$ cannot be targeted in the four-sequence design considered here by simply increasing the number of individuals per cluster-period; the number of clusters must be increased to do so.

\begin{figure}[ht!]
    \centering
    \includegraphics[width=0.7\linewidth]{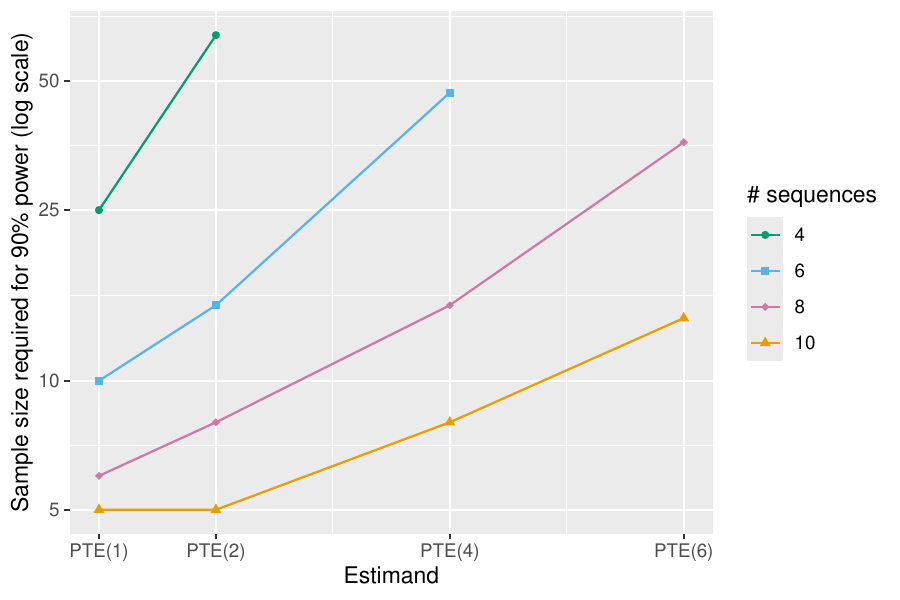}
    \captionsetup{width=0.9\linewidth}
    \caption{Sample size required (number of individuals per cluster-period) for 90\% power with an ETI model as a function of several PTE estimands of interest. Results correspond to a design with 4, 6, 8, or 10 sequences, 5 individuals per cluster-period, a (standardized) effect size of 0.2, an ICC of 0.05, and a CAC of 0.75. Power calculations assume data are analyzed using an ETI model. If no point is plotted for a given combination of estimand and number of sequences then it is impossible to achieve 90\% power for the ETI model with any sample size in that scenario. Lack of smoothness in the 10-sequence trend is due to discreteness in the number of individuals per cluster-period. The Y-axis is displayed on the log scale.}
    \label{fig_SS_ETI_pte}
\end{figure}

Finally, we consider the required sample size for estimation of the point treatment effect at different exposure times (denoted $\text{PTE}(k)$ for $k \in (1,2,...,S)$) using an ETI model. Given patterns observed so far, we expect to see that the required sample size increases as the exposure time of interest increases, and Figure \ref{fig_SS_ETI_pte} confirms that this is indeed the case.

Required sample size increases enormously for estimation of $\text{PTE}(k)$ as $k$ increases. For a design with eight sequences, 6 individuals per cluster-period are required to target $\text{PTE}(1)$, whereas 36 individuals per cluster-period are required to target $\text{PTE}(6)$. This reinforces the message that stepped wedge designs are better for estimating short-term effects than for estimating long-term effects. Intuitively, this trend makes sense, as all sequences in a given study are observed at exposure time $s=1$ but only a single sequence is observed at the largest exposure time.

\subsection{Effects of study design choices: additional time points}\label{ssec_extra_time}

\begin{figure}[ht!]
    \centering
    \includegraphics[width=0.9\linewidth]{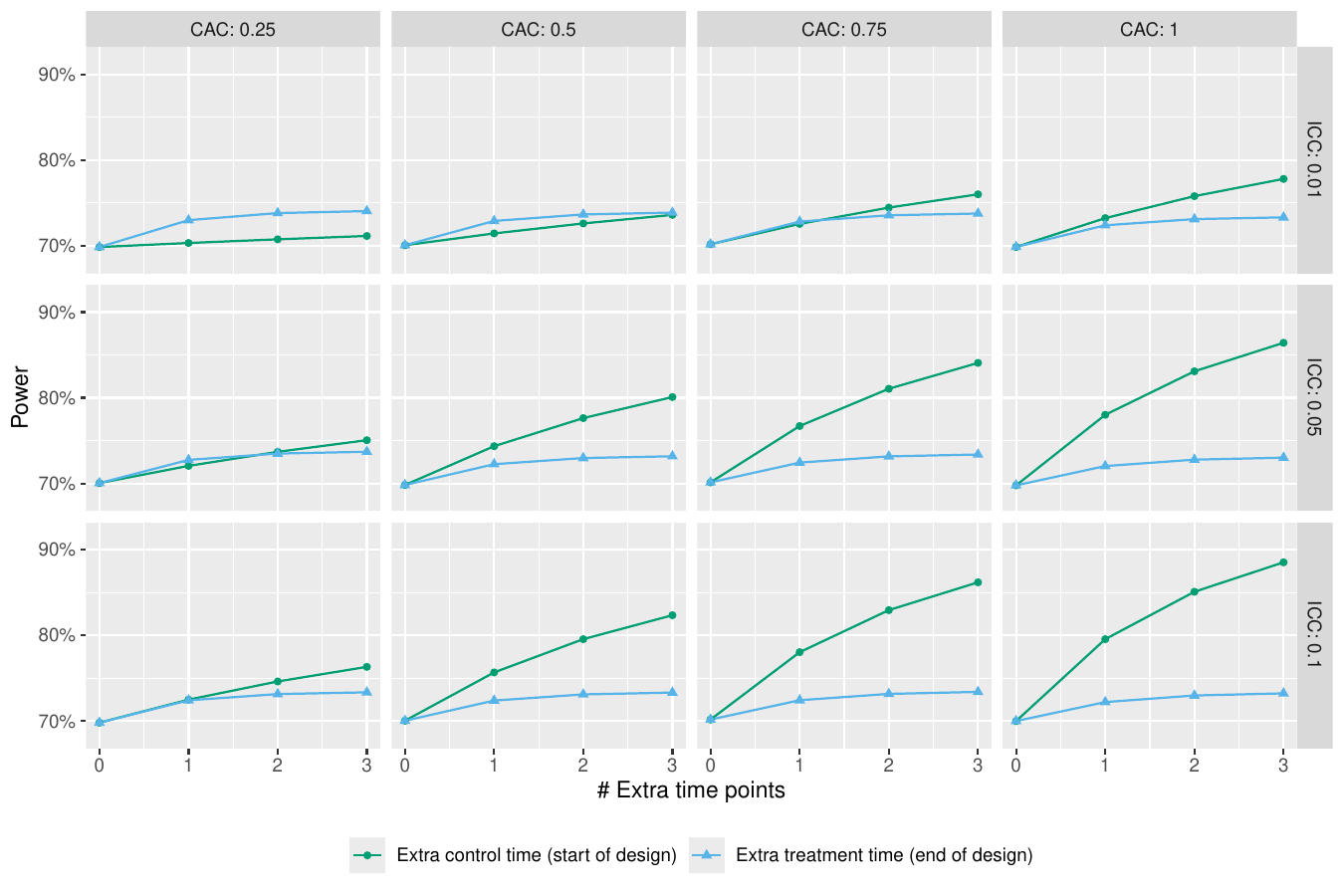}
    \captionsetup{width=0.9\linewidth}
    \caption{Statistical power as a function of additional data collection time points, for estimation of $\text{TATE}(0,6)$, the time-averaged treatment effect over the course of the study, using an ETI model. Results correspond to a design with 6 sequences and 4 clusters per sequence. ICC and CAC values are shown in the facet labels, and the effect size is scaled such that power is 70\% when no extra time is added. Power calculations assume data are analyzed using an ETI model.}
    \label{fig_power_extra_time_tate}
\end{figure}

In this section, we examine the effects of several design choices on power in the context of time-varying treatment effects. We begin by asking the question of whether collecting additional data at either the start or the end of the study can help improve statistical power when interest lies in estimation of the TATE over the course of the study. Figure \ref{fig_power_extra_time_tate} displays power as a function of additional data collection time points, for estimation of $\text{TATE}(0,6)$ using an ETI model (estimated using the \texttt{swCRTdesign} package, as described in section \ref{ssec_methods_formula}). For each combination of ICC and CAC, the effect size is scaled such that the power of the design with no additional time points added is 70\%. The green line displays results for when extra time points are added to the start of the study (i.e., when all clusters are in the control condition) and the blue line displays results for when extra time points are added to the end of the study (i.e., when all clusters are in the treatment condition). Importantly, we do not change the definition of the estimand when considering the addition of time points to the start or end of the study.

Somewhat counterintuitively, adding additional time points to the end of the study leads to only a small gain in power for estimating $\text{TATE}(0,6)$, regardless of the ICC or CAC values. This may be in part due to the fact that vertical treatment-control contrasts are not possible at these later time points, since all clusters are in the treatment state. These results assume the use of an ETI model and that no further assumptions are made about the shape of the effect curve; if additional assumptions are made (e.g., that the effect curve remains constant after a certain exposure time) additional power gains may be possible, although at the risk of bias due to model misspecification.

In contrast, adding additional time points to the start of the study often has a substantial effect on power. For example, when the ICC is 0.05 and the CAC is 1, adding just a single time point to the start of the study increases power by 8\%, and adding three time points increases power by roughly 18\%. As the ICC approaches zero, this power gain disappears. These results are qualitatively similar for designs with different numbers of sequences, numbers of clusters per sequence, and effect sizes. They are also similar for estimation of the TATE over shorter time periods, such as $\text{TATE}(0,4)$ or $\text{TATE}(0,2)$. Furthermore, the effect of adding one time point to the start of the study is identical to the effect of doubling the sample size (individuals per cluster-period) measured in the original baseline period of the study.

Intuitively, the gain in power due to adding time points at the start of the study results from an increase in precision of the estimation of the cluster random effects, as shown analytically in the context of a simple two-sequence design in Appendix \ref{appx_twosequence}. We also observe that the magnitude of this power increase is attenuated with decreasing CAC.

\subsection{Effects of study design choices: the staircase design}

In this section, we study the ``staircase design'' \citep{hooper2014dog,kasza2019information}, a variant of the stepped wedge in which data collection is concentrated immediately before and after the crossover point for each sequence. One reason to consider this design variant is because, if there are time-varying treatment effects, the design implicitly restricts the set of estimands that can be targeted, focusing data collection on only the exposure times that are most efficient to study. To see this, we adopt the notation of \cite{grantham2024staircase} and write $SC(S,K,R_0,R_1)$ to denote a design involving $S$ treatment sequences, $K$ clusters per sequence, $R_0$ periods of data collection in the control state for each sequence, and $R_1$ periods of data collection in the treatment state for each sequence. For a simple $SC(S,K,R_0,1)$ design, for any choice of $(S,K,R_0)$, involving one period of data collection following implementation, the only estimand that can be targeted with respect to exposure time varying treatment effects is the point treatment effect at exposure time one. Thus, if interest lies in time-averaged treatment effects over an extended period or in long-term treatment effects, this design is not appropriate.

\begin{figure}[ht!]
    \centering
    \includegraphics[width=0.9\linewidth]{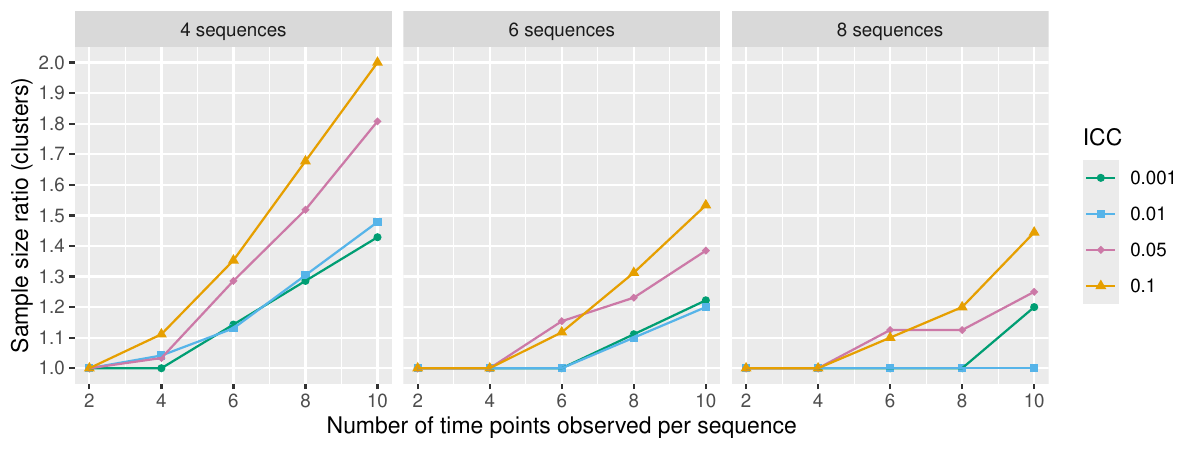}
    \captionsetup{width=0.9\linewidth}
    \caption{Sample size ratio (clusters) required for $\ge$90\% power in a staircase design, shown for designs with 4, 6, or 8 sequences and varying ICC values. The X-axis represents the total number of time points observed (i.e., $R_0+R_1$, with $R_0=R_1$). Results correspond to a design with 4 clusters per sequence, a (standardized) effect size of 0.2, and a CAC of 0.75. Power calculations assume data are generated from an IT model. Lack of smoothness in the trends is due to discreteness in the number of clusters.}
    \label{fig_staircase}
\end{figure}

For a simple staircase design with $R_1=1$, the ETI and IT models are mathematically equivalent. For designs with $R_1>1$, we can examine the cluster SSR using an ETI model targeting $\text{TATE}(0,R_1)$ versus using an IT model. Results are shown in Figure \ref{fig_staircase}, in which the X-axis represents the total number of time points observed (i.e., $R_0+R_1$), where for simplicity we consider designs with $R_0=R_1$. This figure was generated using the \texttt{swCRTdesign} package, as described in section \ref{ssec_methods_formula}.

We can see that in general, the cluster SSR is an increasing function of the ICC and an increasing function of the number of time points observed per sequence; note that the latter is in part due to the fact that the estimand $\text{TATE}(0,R_1)$ changes as the number of time points observed per sequence changes. As in the classic stepped wedge design, there is a price to pay in terms of required sample size when an ETI model is used instead of an IT model in the context of a staircase design.

\subsection{Effects of modeling choices: smoothing the time trends}

\begin{figure}[ht!]
    \centering
    \includegraphics[width=0.9\linewidth]{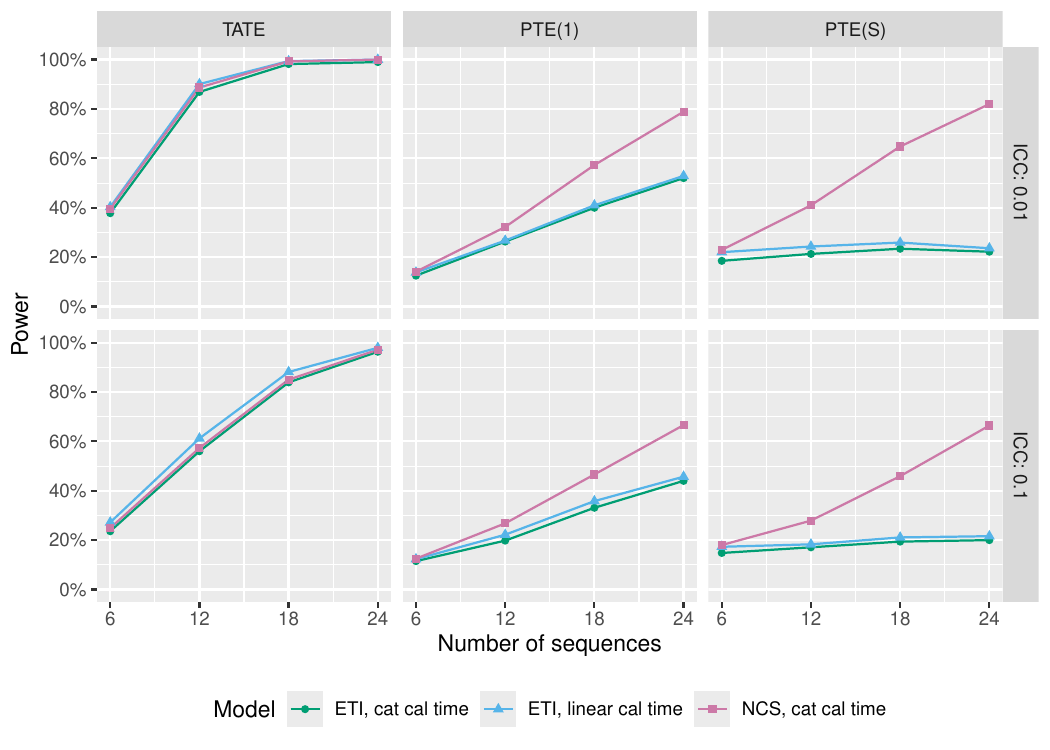}
    \captionsetup{width=0.9\linewidth}
    \caption{Statistical power as a function of number of sequences, shown for three estimands and two ICC values. Colors represent different models, including ETI with categorical calendar time (green), ETI with linear calendar time (blue), and NCS with categorical calendar time (pink). Data were generated according to a model with an effect curve that has an immediate jump from 0 to 0.1 and a subsequent linear increase from 0.1 to 0.3 over the course of the study, 4 clusters per sequence, 5 individuals per cluster-period, and a CAC of 0.75.}
    \label{fig_models}
\end{figure}

In this section, we examine the effects of modeling choices on statistical power. In Figure \ref{fig_models}, we display power for three different mixed models, as a function of number of sequences in the study (for a fixed number of clusters per sequence). Power is estimated via simulation, as described in section \ref{ssec_methods_simulation}. The analysis models include an ETI model with a categorical time trend, an ETI model with a linear calendar time trend (to assess whether power can be gained by placing additional structure on the calendar time trend), and an NCS model with four degrees of freedom and a categorical calendar time trend (to assess whether power can be gained by placing additional structure on the exposure time trend). All three models are correctly specified, since data are generated from an ETI model with a linear calendar time trend, and all models used a random cluster intercept and a random cluster-by-time interaction to model the correlation structure. We generated data for several different effect curves, and show results for a curve in which there is an immediate jump from 0 to 0.1 followed by a linear increase from 0.1 to 0.3 over the course of the study. The six plot facets correspond to two ICC values (0.01 and 0.1) and three different target estimands, the TATE over the course of the study, the point treatment effect $\text{PTE}(1)$ one exposure time point after the start of the intervention, and the point treatment effect $\text{PTE}(S)$ corresponding to the largest exposure time in the study. Note that several lines are jittered slightly for visual clarity.

For estimation of $\text{TATE}(0,S)$, we do not see a substantial gain in power relative to the ETI model with a categorical calendar time trend when we impose additional model structure on the exposure time trend (via the NCS model) or the calendar time trend (via the ETI model with a linear time trend), even though all of these models are correctly specified. For the NCS model, this is consistent with previous results (e.g., Figure 5 of \citealp{kenny2022analysis}); intuitively, the average of smoothed point treatment effect estimates from the NCS model will always be very similar to the average of unsmoothed estimates from the ETI model, and so both the TATE estimate and the variance of the TATE estimate will be similar between these models. Results are similar for an NCS model with a linear calendar time trend; this is not shown on the graph for visual simplicity. For a 24-sequence design, the ETI model with categorical time involves 49 fixed effect parameters (24 exposure time parameters and 25 calendar time parameters), whereas the ETI model with a linear calendar time trend involves 26 fixed effect parameters, the NCS model with categorical calendar time involves 29 fixed effect parameters, and the NCS model with a linear time trend (not shown) involves just 6 fixed effect parameters. Thus, it is somewhat surprising that none of these models do much better than an ETI model with categorical time trend. However, the fact that different calendar time trend parameterizations do not affect power is consistent with previous analytical results related to the immediate treatment model that showed that the variance of the treatment effect estimator is the same regardless of whether a linear time trend or a categorical time trend was used \citep{grantham2020time}. Results are qualitatively similar if data are generated according to different effect curves.

For estimation of $\text{PTE}(1)$ and $\text{PTE}(S)$, we do see a gain in power associated with imposing additional structure on the exposure time trend via the NCS model. This trend is especially pronounced for estimation of the long-term PTE, and the gain in power is larger for designs with greater numbers of sequences. However, we do not see any gain in power whatsoever associated with imposing additional structure on the calendar time trend. Results are qualitatively similar for other CAC values.

\subsection{Effects of modeling choices: including washout periods}

\begin{figure}[ht!]
    \centering
    \includegraphics[width=0.9\linewidth]{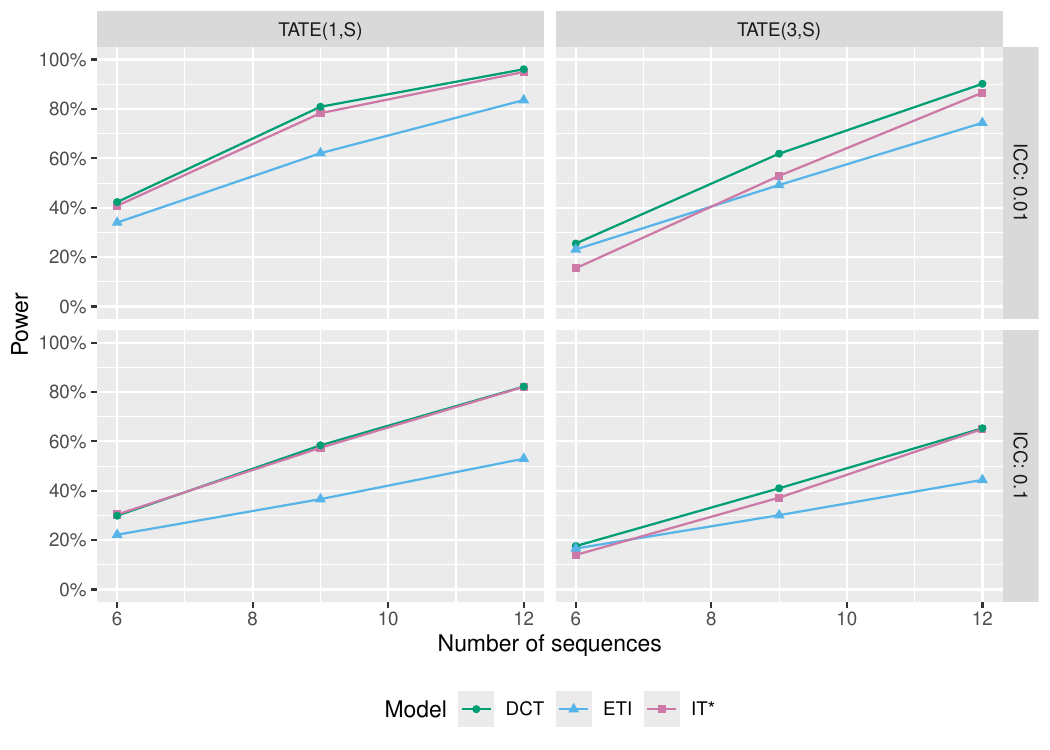}
    \captionsetup{width=0.9\linewidth}
    \caption{Statistical power as a function of number of sequences, shown for two estimands and two ICC values. Colors represent different models, including an ETI model (blue), a modified IT model that drops data corresponding to the washout period (IT*; pink), and a delayed constant treatment model (DCT; green), all with categorical calendar time indicators. Data were generated according to a DCT model with an immediate treatment effect of $\delta=0.2$, 4 clusters per sequence, 5 individuals per cluster-period, and a CAC of 0.75.}
    \label{fig_washout}
\end{figure}

Next, we consider how power compares between several models that can be used to analyze a stepped wedge dataset if it is assumed that the treatment effect is constant, but only after a washout period passes. It is common in many fields to consider washout periods \citep{wils2024washout,harvey2021modernizing}, and in the context of cluster randomized trials, it is sometimes suggested that data corresponding to the washout period is discarded or not collected in the first place \citep{caille2024practical}. However, other approaches are possible, and in this section, we compare the performance of three models: a modified IT model that drops data corresponding to the washout period, an ETI model, and the delayed constant treatment (DCT) model described in section \ref{ssec_models}. The DCT model allows for a flexible exposure time trend during the washout period but assumes that the treatment effect is constant following this washout period. Power is estimated via simulation, using the same data-generating mechanism described above, but with an immediate treatment effect of 0.2. Results are displayed in Figure \ref{fig_washout}, with the left two panels showing results for estimating $\text{TATE}(1,S)$ (i.e., a washout period of length 1) and the right two panels showing results for estimating $\text{TATE}(3,S)$ (i.e., a washout period of length 3).

As expected, the ETI model has the lowest level of power across all scenarios. In some scenarios, the DCT model displays slightly higher power than the modified IT model, and it appears that the power gain is more prominent for lower ICC values and designs with smaller numbers of sequences. Results are qualitatively similar for other CAC values.

\section{Discussion}\label{sec_discussion}

In this work, we characterized factors related to estimands, study design, and modeling that influence the power of stepped wedge cluster randomized trials (SW-CRTs) when treatment effects vary with exposure time. There are several practical takeaways for researchers designing SW-CRTs, and it is critical to think about these points at the planning stage, since they may have an enormous influence on the sample size required for a given trial.

The choice of estimand has substantial influence on statistical power. For a standard SW-CRT in which the IT model is correct, the sample size (clusters per sequence or individuals per cluster-period) must be increased substantially, often by a factor of 2 to 3, to maintain 90\% power when switching from an IT model to an ETI model (targeting the TATE over the course of the study), unless the ICC is very low; we refer to this inflation factor as the sample size ratio (SSR). We emphasize that this is for a setting in which the IT model is correctly specified; if it is not, estimation using an IT model may be severely biased and inference highly incorrect. The SSR is lower for ``short-term TATE'' estimands ($\text{TATE}(0,S-k)$ for $k>0$) and higher for ``long-term TATE'' estimands ($\text{TATE}(k,S)$ for $k>0$). In general, when using an ETI model, the required sample size is much lower for estimating short-term effects (e.g., PTE(1) or TATE(1,3) for an eight-sequence design) relative to long-term effects (e.g., PTE(8) or TATE(6,8)). For example, in an eight-sequence design that requires 10 individuals per cluster-period to target $\text{TATE}(0,8)$, the sample size can be cut in half to 5 individuals if one instead targets $\text{TATE}(0,5)$ but must be doubled to 20 individuals to target $\text{TATE}(3,8)$. Similarly, for a design involving eight sequences requiring roughly 6 individuals per cluster-period to target $\text{PTE}(1)$, the sample size must be increased to 36 individuals per cluster-period to instead target $\text{PTE}(6)$. Intuitively these results make sense, as all sequences are observed at exposure time $1$ whereas only one sequence is observed at exposure time $S$.

For a given estimand, many design choices may potentially affect power; in this work, we examined the impact of additional data collection time points and the impact of using a staircase design. Intuitively, one might guess that collecting additional data at the end of the study would improve power for targeting TATE estimands, since this results in more observations corresponding to higher exposure times. Surprisingly, our results show that this gain is minimal when an ETI model is used. In contrast, adding additional data collection time points to the start of the study or increasing the sample size (individuals per cluster-period) at baseline can result in a sizable gain in power. For example, with a six-sequence, for an ICC value of 0.05 and a CAC of 1, adding a single time point to the start of the study (or doubling the baseline sample size) increases power by roughly 8\%, and adding three time points (or quadrupling the baseline sample size) increases power by roughly 18\%. This power gain is larger for higher ICC and CAC values, and as the ICC approaches zero, this power gain disappears. Intuitively, these additional data collection points at the start of the study help improve the estimation of cluster random effects, which in turn improves the precision of the TATE estimator (see Appendix \ref{appx_twosequence} for an analytic argument that gives some intuition for why this occurs). Next, when considering the staircase design, it should be understood that this design inherently restricts the set of estimands that can be targeted. For example, in the ``classic'' staircase design with two time points of data collection, one before the intervention and one after the intervention, the only estimand that can be targeted is the PTE at exposure time one. Using an ETI model with a staircase design involving more than two time points of data collection following the intervention requires a larger sample size, especially for designs with higher ICCs and many data collection time points following the intervention.

Using more restrictive models for the calendar time trend and/or the exposure time trend may help in terms of power, but only in certain situations. Surprisingly, using the correct parametric form for the calendar time trend instead of a categorical trend results in virtually no gain in power for any estimand considered (the TATE over the course of the study, the PTE at exposure time 1, or the PTE at the largest exposure time), while risking bias due to incorrect specification. Modeling the exposure time trend using a natural cubic spline did not increase power for the estimation of the TATE over the course of the study, but resulted in substantially increased power if interest lies in the PTE (at any time point). If it can be assumed that the treatment effect is constant after a washout period passes, we recommend the use of the ``delayed constant treatment'' (DCT) model defined in section \ref{ssec_models}, which includes data corresponding to the washout period in the model but imposes no structure on the shape of the effect curve during this period. This has the dual advantage of slightly increasing power (particularly in designs with smaller numbers of sequences and with lower ICC values) and allowing for the effect curve during the washout period to be estimated.

Whenever possible, increasing the number of clusters will virtually always be a better strategy than increasing the number of individuals per cluster-period (with respect to precision and validity of treatment effect estimation) for several reasons. First, the gain in power from increasing the number of clusters will usually be higher than the gain in power from increasing the number of individuals sampled per cluster-period by the same factor. As demonstrated, it is sometimes not possible to achieve 90\% power by increasing the number of individuals per cluster-period along, whereas it will always be possible to do so by increasing the number of clusters per sequence. Second, increasing the number of clusters will improve estimation of cluster variance components. Third, and perhaps most importantly, increasing the number of clusters improves the likelihood of achieving balance with respect to unmeasured cluster-level confounding variables.

There are a number of immediate extensions to this work that would be of use to trial designers and analysts. Although we examined the impact of several design features, including the addition of extra time points to the start or end of the study and use of the staircase design, future research is needed to determine optimal design for different estimands. In particular, recent work that examines the impact of incomplete designs, unequal allocation of observations to cluster-period cells, and unequal allocation of clusters to sequences, such as \cite{thompson2017optimal}, \cite{hooper2020hunt}, and \cite{rezaei2023impact}, must be re-examined in the context of time-varying treatment effect estimands, as existing results assume an immediate treatment effect. Furthermore, work is needed to determine which designs are optimal for estimating longer-term effect measures, since SW-CRTs (including staircase designs) are clearly not a good choice for targeting these estimands; it could be the case that alternative allocation patterns make it more feasible to estimate long-term effects. Similarly, it is worth revisiting stepped wedge methodological research examining model misspecification \citep{voldal2022model,thompson2017bias,ouyang2024maintaining}, and robust estimation and inference for treatment effects \citep{hughes2020robust,thompson2018robust,kennedy2020novel}. We also restricted our analysis to the use of linear mixed models with continuous outcomes and fairly simple correlation structures. Future analyses can look at sensitivity of results to alternative analysis models (such as GEE models or mixed models with different robust variance estimators), binary/count outcomes, and more complex correlation structures. As mentioned in section \ref{sec_intro}, \cite{hughes2024sample} demonstrated that under an ETI mixed model with an identity link function, the variance of TATE and PTE estimators (that can be expressed as linear combinations of the point treatment effect estimators) does not depend on the shape of the effect curve, but only on the summary being targeted; however, future research can examine the influence of the shape of the effect curve on power in settings with a nonlinear link function. Finally, although we focused on settings in which the treatment effect varies as a function of exposure time, analogous results for settings in which the treatment effect varies as a function of calendar time, as studied by \cite{wang2024achieve} and \cite{lee2024analysis}, would be a useful contribution.

\section{Conclusions}

Factors related to the choice of estimand, study design details, and analysis model selection can have enormous influence on statistical power in stepped wedge cluster randomized trials. Researchers should think proactively about each of these factors when planning a trial.

\section{List of abbreviations}

\quad\begin{tabular}{ l l }
SW-CRT & Stepped wedge cluster randomized trial \\
IT & Immediate treatment \\
ETI & Exposure time indicator \\
TATE & Time-averaged treatment effect \\
PTE & Point treatment effect \\
ICC & Intraclass correlation coefficient \\
CAC & Cluster autocorrelation coefficient \\
DCT & Delayed constant treatment \\
NCS & Natural cubic spline \\
SSR & Sample size ratio \\
GEE & Generalized estimating equations
\end{tabular}

\section{Declarations}

\subsection{Ethics approval and consent to participate}

Not applicable.

\subsection{Consent for publication}

Not applicable.

\subsection{Availability of data and materials}

Code to reproduce all analyses and simulations is available at \url{https://github.com/Avi-Kenny/SW-Power}.

\subsection{Competing interests}

The authors declare that they have no competing interests.

\subsection{Funding}

Research reported in this publication was supported by the National Institute Of Allergy And Infectious Diseases of the National Institutes of Health under Award Number R37AI029168. The content is solely the responsibility of the authors and does not necessarily represent the official views of the National Institutes of Health.

\subsection{Authors' contributions}

AK wrote all analysis and simulation code. All authors helped to conceptualize the study, provided critical scientific input, read the final manuscript, and approve of all content.

\subsection{Acknowledgements}

Not applicable.

\appendix

\section*{References}

{\setstretch{1.0}
\renewcommand\refname{\vskip -1cm}
\bibliography{refs.bib}
}

\section{Analytic results for two-sequence design}\label{appx_twosequence}

In this section, we analytically examine the forms of the point treatment effect estimators $\hat{\delta}_1$ and $\hat{\delta}_2$ resulting from an ETI model in the context of several simple designs, in order to illustrate the intuition behind the results shown in Figure \ref{fig_power_extra_time_tate}. First, consider the ``base design'' given in Table \ref{appx_design_base}, a simple two-sequence/two-cluster stepped wedge design in which the $Y_{ij}$ terms represent the observed cluster-period means.

\begin{table}[ht!]
    \centering
    \begin{tabular}{|l|l|l|} \hline
    $Y_{11}$ & \LG$Y_{12}$ & \LG$Y_{13}$ \\ \hline
    $Y_{21}$ & $Y_{22}$ & \LG$Y_{23}$ \\ \hline
    \end{tabular}
    \captionsetup{width=0.5\textwidth}
    \caption{The ``base design'', a standard stepped wedge design with two sequences. Control periods shown with white cells and treatment periods shown with grey cells.}
    \label{appx_design_base}
\end{table}

A modification of the base design in which each cluster is observed for one additional treatment period at the end of the study (with the new observations denoted $Y_{14}$ and $Y_{24}$) is given in Table \ref{appx_design_add1t}.

\begin{table}[ht!]
    \centering
    \begin{tabular}{|l|l|l|l|} \hline
    $Y_{11}$ & \LG$Y_{12}$ & \LG$Y_{13}$ & \LG$Y_{14}$ \\ \hline
    $Y_{21}$ & $Y_{22}$ & \LG$Y_{23}$ & \LG$Y_{24}$ \\ \hline
    \end{tabular}
    \captionsetup{width=0.5\textwidth}
    \caption{The ``Add-1T design'', equivalent to the base design, but with one additional treatment period added to the end of the study. Control periods shown with white cells and treatment periods shown with grey cells.}
    \label{appx_design_add1t}
\end{table}

An alternative modification of the base design in which each cluster is observed for one additional control period at the start of the study (with the new observations denoted $Y_{10}$ and $Y_{20}$) is given in Table \ref{appx_design_add1c}.

\begin{table}[ht!]
    \centering
    \begin{tabular}{|l|l|l|l|} \hline
    $Y_{10}$ & $Y_{11}$ & \LG$Y_{12}$ & \LG$Y_{13}$ \\ \hline
    $Y_{20}$ & $Y_{21}$ & $Y_{22}$ & \LG$Y_{23}$ \\ \hline
    \end{tabular}
    \captionsetup{width=0.5\textwidth}
    \caption{The ``Add-1C design'', equivalent to the base design, but with one additional control period added to the start of the study. Control periods shown with white cells and treatment periods shown with grey cells.}
    \label{appx_design_add1c}
\end{table}

Assume that all three designs are analyzed with an ETI model, specifically a linear mixed model with a cluster random intercept and point treatment effect parameters $\delta_1$ and $\delta_2$ (and, in the case of the add-1T design, $\delta_3$), as well as time parameters $\beta_1$, $\beta_2$, and so on. For the base design, it can be shown that the resulting estimators $\hat{\delta}_1$ and $\hat{\delta}_2$ of the point treatment effect parameters $\delta_1$ and $\delta_2$ are given by

\begin{align}
\begin{split}\label{delta_hats_base}
    \hat{\delta}_1 &= (Y_{12}-Y_{22}) - \phi(Y_{11}-Y_{21}) \,, \\
    \hat{\delta}_2 &= (Y_{13}-Y_{23}) + \hat{\delta}_1 - \phi(Y_{11}-Y_{21}) \,,
\end{split}
\end{align}

\noindent where $\phi = \tau^2/(\tau^2+\sigma^2/K)$, $\tau^2$ is the cluster-level variance, $\sigma^2$ is the individual-level variance, and $K$ is the number of individuals per cluster-period. The estimator $\hat{\delta}_1$ can be seen as the sum of two terms: (a) the vertical difference at time 2 and (b) the ``cluster difference'' (i.e., the baseline difference in means between the two clusters, scaled by $\phi$). When $\phi=1$, $\hat{\delta}_1$ can also be seen as equivalent to a difference-in-differences estimator. The estimator $\hat{\delta}_2$ can be seen as a sum of three terms: (a) the vertical difference at time three, (b) the estimator $\hat{\delta}_1$, and (c) the cluster difference.

For the add-1T design, in which an additional time point is added to the end of the study, it turns out that the estimators of $\delta_1$ and $\delta_2$ are identical to those given in \eqref{delta_hats_base}. Intuitively, this is because the data point $Y_{14}$ is used by the model to estimate $\delta_3$, the point treatment effect at exposure time 3, and the data point $Y_{24}$ is used to estimate the calendar time effect $\beta_4$. In other words, an ETI model for the add-1T design involves two additional data points and two additional parameters, so no additional information is available to improve estimation of $\delta_1$ or $\delta_2$. While this argument is specific to the two-sequence design, it sheds some light on why we do not observe a large gain in power when adding a time point to the end of the study.

For the add-1C design, in which an additional time point is added to the start of the study, the estimators $\hat{\delta}_1^*$ and $\hat{\delta}_2^*$ of the point treatment effect parameters $\delta_1$ and $\delta_2$ are given by

\begin{align}
\begin{split}\label{delta_hats_add1c}
    \hat{\delta}_1^* &= (Y_{12}-Y_{22})
    - \frac{\phi}{1+\phi} \left\{ (Y_{10}+Y_{11}) - (Y_{20}+Y_{21}) \right\} \,, \\
    \hat{\delta}_2^* &= (Y_{13}-Y_{23}) + \hat{\delta}_1
    - \frac{\phi}{1+\phi} \left\{ (Y_{10}+Y_{11}) - (Y_{20}+Y_{21}) \right\} \,.
\end{split}
\end{align}

Examining the estimators in \eqref{delta_hats_add1c}, we see that they are similar to those given in \eqref{delta_hats_base}, but with a different ``cluster difference'' term. Specifically, the cluster difference terms involve a factor $\frac{\phi}{1+\phi}$ (instead of just $\phi$), and are calculated with the ``pooled'' data from time point 0 (i.e., the added time point) and time point 1. This results in an increase in precision in estimating the cluster difference, which in turn leads to increased precision for estimating $\delta_1$ and $\delta_2$. Furthermore, these expressions illustrate why the precision gain is more prominent for data-generating mechanisms involving higher ICCs, since the factor $\phi$ will approach zero as the ICC approaches zero.

For a design involving three or more sequences, the expressions for the point treatment effect estimators become much more complicated in form and thus more difficult to analyze. However, it is reasonable to conjecture that the intuition is similar, in the sense that the precision gain resulting from additional time points added to the start of the study is largely due to the increased efficiency in estimating cluster differences.

\end{document}